\documentclass[eps,jmp,amsmath,amssymb]{revtex4-1}
\usepackage{epsfig}
\usepackage{epstopdf}
\usepackage{slashbox}
\usepackage{textcomp}
\usepackage{graphicx}
\usepackage{dcolumn}
\usepackage{bm}
\usepackage{hyperref}

\usepackage{caption}
\usepackage{subcaption}
\begin{document}
\preprint{AIP/123-QED}
\title[Sample title]{Calculation of Efficiency and Power Output by Considering Different Realistic Prospects for Recovering Heat from Automobile using Thermoelectric Generator}
\author{Kumar Gaurav}
\email{kumargauravmenit@gmail.com}
\altaffiliation{School of Engineering, Indian Institute of Technology Mandi, Kamand, Himachal Pradesh, India, 175005}
\author{Shashank Sisodia}
\author{Sudhir K. Pandey}
\affiliation{School of Engineering, Indian Institute of Technology Mandi, Kamand, Himachal Pradesh, India, 175005}
\date{\today}
\begin{abstract}
Development of method to calculate the efficiency and the methodology to enhance the efficiency is not the ultimate goal. Mainly the calculated result obtained by using mathematical relationship and physical law should be applied in day to day life applications for the benefit of society. In this work, we are developing the theoretical prototype and improving the technique for installing the Thermoelectric Generator (TEG) set up in automobiles. We have applied the methodology for enhancement in efficiency of TEG and make it economical and user friendly. We have considered a linear curve fit from a zigzag curve of reported mass flow rate and temperature variation at the hot gas inlet. Accordingly the temperature of the coolant is also varied linearly at inlet from 300 K to 320 K and corresponding the mass flow rate of coolant. Circular fin is installed around each circular layer of Thermoelectric Materials (TEM) after the water jacket. The heat loss through each fin is calculated as 24 W. Energy balance is done at each and every TEM and correspondingly calculated the amount of power transferred through each segment of TEG as 32 W. We have calculated the length of TEM sample for attaining the respective temperature range for the hybrid of $Bi_2Te_3$ and $TiO_{1.1}$. This calculation is done by considering the compatibility factor derived as $s =\frac{\sqrt{1+ ZT}-1}{\alpha T}$ which is a function of only intrinsic material properties and temperature and is represented by a ratio of current density to conduction heat flux. The length obtained for this particular combination is $\sim$8 mm. To this end, we have reported the efficiency with respect to mass flow rate of hot flue gas from automobile for different layers of TEG for the above mentioned combination. Here, we have explored the possibility of installing a number of different layers TEG module which can be installed throughout the lateral surface area of exhaust chamber. Thermal mismatching criteria are also discussed at the adjoining surface of TEM because of high temperature. To maintain the thermal expansion or contraction of TEM, spring and bolt arrangement is provided, which is fixed over the aluminium oxide ceramic substrate. Ideal power output by first layer is also calculated by temperature dependent materials parameter, which is 100 W and plotted in graph showing the change in power output with variation in temperature of sink. For automobile, if temperature of source is considered as 800 K, so for the temperature range of 300 K to 800 K is obtained as 58 W.\\
\end{abstract}
\bigskip
\maketitle
\thispagestyle{plain}
\pagestyle{plain}
\section{INTRODUCTION}
Global warming is expanding their mouth like monster, which is a huge challenge for living organisms to survive on earth. Since, for survival and fulfilling basic necessity of human, the power is as necessary as others. The consumption of power leads to increase in entropy and consequently raising the temperature of the universe. Either, the human beings have to reduce their day to day increasing power consuming luxurious amenities or to think of alternate way of extracting the waste heat from some heat source. Among the different available methods for extracting the useful energy from waste heat, the TEG can be a good option to explore\cite{Murakami, Bell, Yu, Wu, Kraemer}. The reason behind the selection of TEG as well as the different available heat resources for installing TEG is already discussed in our previous article\cite{Gaurav}. Selection of particular material for a specific application like stove, solar\cite{Chen, Omer} is also discussed with proper justification\cite{Gaurav}. Considering idealistic situation in different probable locations like an automobile or the steel industry, we have calculated the efficiency of TEG with different temperature condition at hot exhaust with varying hot or cold end temperature.

In spite of all the discussed problems in our previous article\cite{Gaurav}, we have encountered lots of different challenges which are to be resolved. The calculation of efficiency as well as materials selection will be justified only when the setup can be installed in an automobile by considering different possible aspects. This comprises of maintaining constant heat sink temperature by flowing coolant. The temperature of coolant is itself a problem because after continuous flow of water the temperature of coolant will rise up. The constant heating problem of coolant is removed by installing double direction flow of coolant. Another aspect is to remove the extra heat from the coolant, which is done by using fins. Our aim is to extract as much amount of energy from the flowing hot exhaust flue gas through the cylindrical chamber. The next challenge is to design the length of the chamber through which hot flue gas will flow. Obtaining the proper length of TEM to attain the corresponding temperature difference across the sample by considering the variable thermal conductivity parameter. So, we have to think for optimization of different aspects with respect to specific parameter of setup for installation the module in an automobile for efficiency calculation.

For installation of the TEG in real life application, we have reviewed few of the articles. Orr et al.\cite{Orr} used heat pipes in addition to TEM module. The area of flow of coolant on the hexagonal mesh is very less which will remove very less heat. Subsequently the temperature difference across the sample is very low, which gives low efficiency. The dimension of the system is large and extra added weight will increase the inertia. Korzhuev\cite{Korz} considered the IC Engine and TEG as single thermodynamics system and then defined their economic feasibility. Their observation was conflict between power output and back pressure caused by adding TEG. Due to the addition of TEG, induced back pressure caused a drop in efficiency. The research gave us direction for carefully installing the TEG setup without disturbing the efficiency of internal combustion engine with new developed back pressure. Liu et al.\cite{Liu} performed a case study on the installation of the TEM module on different location of the exhaust pipe of a car. The result indicated that TEG module should be installed between catalytic converter and muffler in order to obtain a uniform high surface temperature of the exhaust pipe. They have not considered about design and material aspects. Madhav\cite{Madhav} studied $Bi_2Te_3$ based TEM and assumed perfect cooling of the exhaust gas in the heat-exchanger. This is in reality not feasible to assume perfect cooling of the exhaust gas because that needs a large setup which will increase efficiency, but correspondingly it will increase the inertia of the vehicle. Wang et al.\cite{Wang} theoretically studied the effect of various factors, including exhaust gas mass flow rate, mass flow rate of different cooling fluids, convective heat transfer coefficient and the height of thermoelectric material. They studied about the height of TEM module, which was based on the assumption of constant thermal conductivity of the p-type and n-type TEM but we know that $\kappa$ is material's parameter which will vary with temperature. Jorge et al.\cite{Jorge} studied the importance of using internal fins in order to maximize heat transfer and turbulence in flue gas results in the increase of energy output. They didn't explain the chances of improvement in the TEG setup because of external coolant supply. Sumeet et al.\cite{Sumeet} studied 1-D heat flux and temperature variation along the thermoelectric legs. The value of TEG leg height was calculated by considering variable material parameter, but they have not used hybrid technology to enhance the efficiency. Champier et al.\cite{Champ} calculated the efficiency of TEG installed in cook stove\cite{Champier}. This gives very lesser power output.

We have discussed the efficiency obtained by installing TEG by considering different possible aspects to maximize the heat sink\cite{Mayer, Yazawa, Lawrence}, because higher efficiency can be obtained in the lower temperature range. The increase in efficiency is obtained because of two reasons, first the Carnot efficiency gives higher efficiency in the low temperature region when comparison is done between two fixed temperature difference and the second reason is that normally the materials we have selected is having a good figure of merit in the lower temperature range. The thermodynamic setup usually works properly in the lower temperature range because at higher temperature region, it creates more wear and tear problem. Another aspect should be noted that the setup will create more entropy in higher temperature. Inside of hot exhaust chamber we made some extruded surface to make turbulence for extracting more heat. We have discussed the calculated efficiency by studying all individual aspects and plotted them for different layers of TEG. We have provided a lot of space for exploring the chances of increasing the layer and correspondingly the power output.

Here, we have addressed the above problem in a simple and precise manner. We have considered the TEG for installation in an automobile, for which, there are some reported results of mass flow rate and corresponding temperature. For example taking Honda city car having 1343 cc engine normally tested to produce 70-80 KW power under some conditions\cite{Jahirul}. In our case the heat extracted by coolant is 5.260 kW during the flow of coolant through the exhaust channel. The heat released by each external fin to the atmosphere is 24 W. TEG module installed in each edge will remove $\sim$30 W. The length obtained for the flue gas chamber where the module is installed is 120 mm. The optimized Reynolds number of turbulent flows is 4132. Accordingly the power required to continue the flow of coolant is 20 W. Since, the vehicle has to travel in different conditions of road and engine, hence the mass flow rate of hot flue gas will change. So, considering all these varying parameters, we have calculated the efficiency with respect to the mass flow rate of hot flowing gas. Ideal power output through the TEG considering the temperature dependent materials parameter is also calculated and plotted in graph.
\section{SETUP PROTOTYPE AND ASSUMPTIONS}
We have considered following aspects during designing of TEG setup which is to be installed in automobiles. 1) We have analyzed the system which is to be installed in the middle of automobile's exhaust pipe and to keep at that position we can weld the TEG at that section. Two aspects we have taken into account, first is there should not be large back pressure, otherwise this will decrease the automobile's engine efficiency. Secondly, TEG module should not be to far from automotive exhaust otherwise there is a huge loss of energy before reaching to the TEG. 2) For the prototype we have assumed, the TEM is installed over a regular octagon. The reason behind selection of such edge like structure is to provide a perfect contact between the heat source and TEM to allow more passage of heat. We have avoided the circular cylinder on which TEG could be installed because it need circular shape of TEG, whose cross sectional area will change with radial distance and consequently the application of Fourier law will become critical. Using different methods to solve the problem will gives rise to error in calculation. 3) Number of layers for installing TEG is assumed to be five for a simple prototype calculation but we will have plenty of space for adjusting the large number of circular layers of TEG. 4) The length of the chamber on which TEG is being installed was decided by many factors like the type of TEM materials which is going to be used, temperature of exhaust gas, the amount of heat extracted by coolant as well as by fin. 5) Circular fin was designed for providing a large contact area with sample to extract a huge amount of energy, which is to be released into the atmosphere. 6) Some extended metallic part was provided inside the chamber of hot exhaust pipe so as to increase the turbulence and more surface area to increase convection and conduction towards the TEM sample. 7) Coolant passes through the rectangular cross sectional chamber, which is having perfect contact with sample to extract more heat. Here, it is important to note that the surface of coolant cross section is convex to some extent from one side, but for calculation part is concerned, we have assumed flat surface. 8) There should not be any leakage of hot exhaust gas from the regular octagonal chamber. This is achieved by providing a concentric lateral octagonal surface parallel to the hot side of sample throughout the entire length except at the sample site. We have taken into consideration about the following aspects like the thickness of a regular octagon which is very thin and it should be metallic as well.
\begin{figure}[h]
\centering
\includegraphics[width=5.5in]{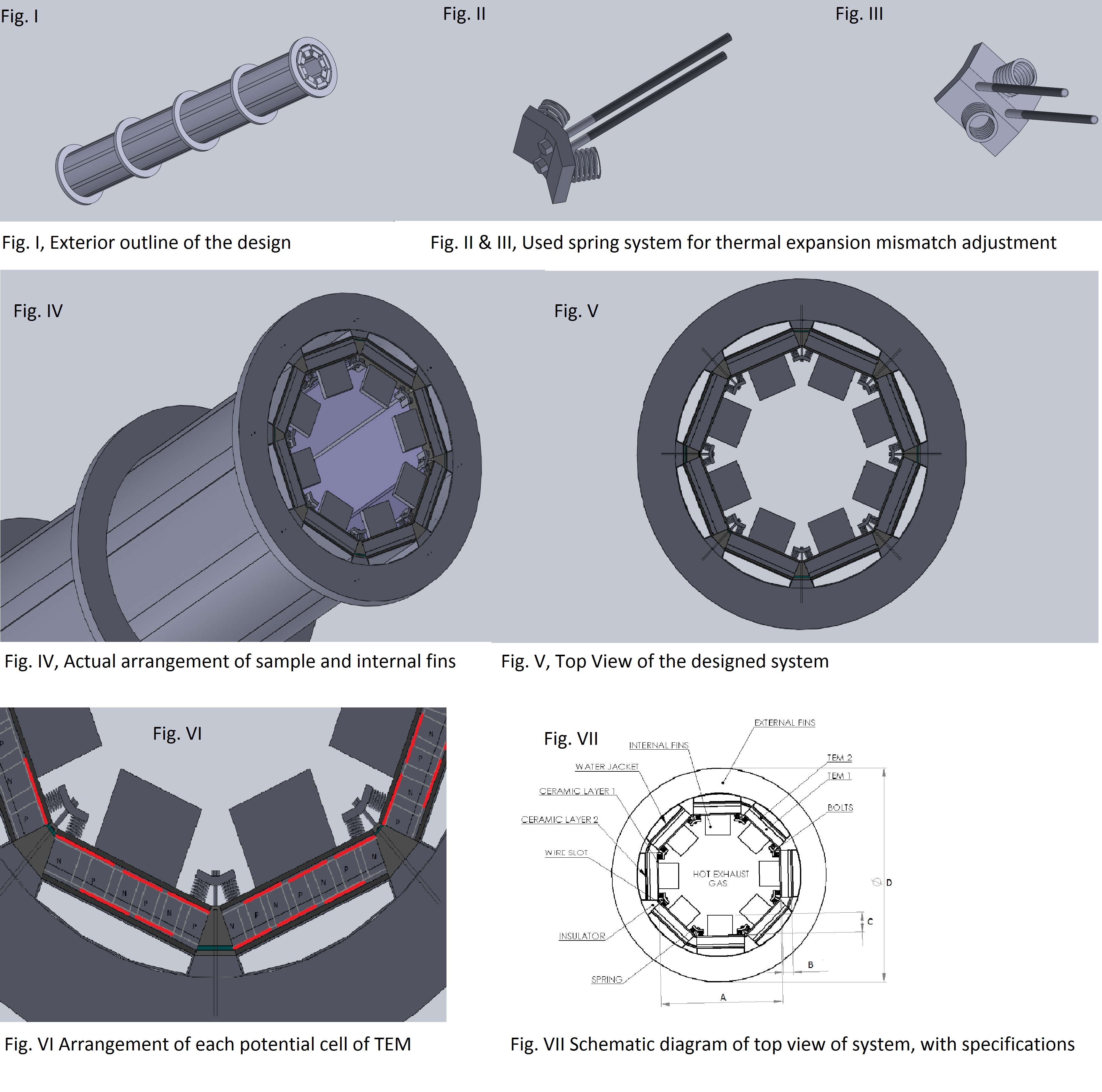}
\caption{Schematic of the design, which is considered for solving the problem}
\end{figure}
In the Fig.(1) we are trying to show different assembly and schematic design. Fig. I show the side view of the setup. Fig. II and Fig. III shows the spring arrangement for longitudinal thermal distortion, during heating or cooling there is a chance of thermal contraction or expansion, so to avoid the mismatch or distortion in sample we have provided spring. Fig. IV, V, VI is used to show the top view of the design and zoomed picture. Fig. VII shows the schematic with the nomenclature of the top view of the design. Here, A, B, C, and D as shown in figure are the diameter of circular chamber, B is thickness of sample and C is depth of internal extruded surface, D is outer diameter of external fin, respectively. The circulation of coolant is done on the basis of maintaining the same temperature in each cycle because the water is flown between two water tanks and the temperature of water is maintained in equilibrium with atmosphere so every time we can assume the inlet temperature of coolant is ambient temperature. The TEM sample is installed over the aluminium oxide ceramic substrate and this substrate is extended throughout the cylinder wherever the TEM sample will be fixed. Rest part will be covered by thermally and electrically insulated materials to prevent the hot flue gas to pass within the vacant space around the sample. The reason behind the taking aluminium oxide ceramic substrate is that it is having thermally conducting and electrically insulating behavior, so this will easily pass the heat energy but electrically prevent from short circuiting.
\section{METHODOLOGY}
Calculation of the length of the cylindrical chamber follows the energy balance. Here one important point is that we can't apply the formula for heat-exchanger because it is not an isolated system, here energy is being lost to the atmosphere through the fin. So, we analytically calculated the length of the hot flue gas chamber with energy balancing. To avoid more complexities, we have taken five layers on our setup on which TEG is being installed. The temperature of each hot end of TEG is decided by the temperature profile of exhaust flue gas passing through the chamber. By applying the energy balance in radial direction, we have formulated the heat passing through the TEM sample. Here, it is mandatory to mention that the potential difference generated across the sample is dependent on temperature difference. So, for large sample size, we will have same potential difference as for the small size sample. Keeping in mind about this aspect we have taken large number of small-small cell modules. Here, cell module is defined as the combination of n-type and p-type TEM forming an electrical series. The separation between each cell module is ideally taken as infinitesimally small. One necessary assumption we have taken to make the system ideal, the entire calculation is done on the basis of neglecting the inter-module separation. Now, the heat loss through the circular fin attached around the water jacket is calculated. Heat flow through the circular fin by conduction and convection follows the second order differential equation as in Eq.(1).
\begin{equation}
\centering
\frac{d^{2}\theta}{dr^{2}}+\frac{1}{r}\frac{d\theta}{d r}-\frac{h}{\kappa B}\theta =0
\end{equation}
Here, r is the radial distance from center of the disc, $\kappa$ is thermal conductivity of fin material, $h$ is convective heat transfer coefficient and $\theta=\frac{T-T_a}{T_w-T_a}$. The aim behind the installation of the fin was to increase the surface area for large heat loss to the atmosphere. The above differential equation will be solved by using modified Bessel function of the first kind and second kind for V = 0,1,2,3,4.., here $I_v (x)$ is used to denote the kind, where v is order (real number) and x is a complex number. Boundary conditions assumed for solving the problem is the rate of heat flow at the farthest point of fin surface from the axis is zero. This means the temperature at the edge of the fin should be same as the temperature of ambient and temperature at distance r from the longitudinal axis is same as the surface temperature of the hot exhaust pipe. The proposed diagram of the design for fin around the cylinder is plotted in the figure. 
\begin{figure}[h]
\centering
\includegraphics[width=3in]{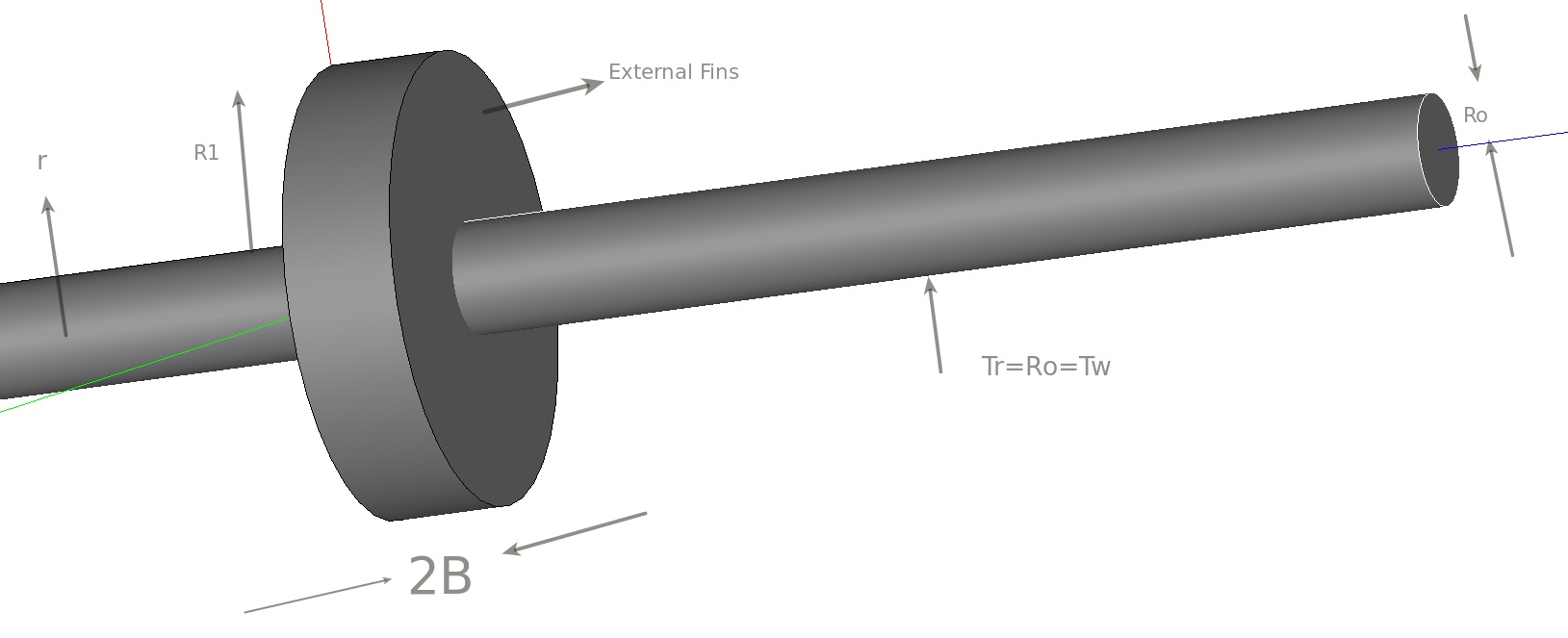}
\caption{The design prospects used for applying a modified Bessel function}
\end{figure}
The generalized solution\cite{heat} for the differential Eq.(1) is shown in Eq.(2), which is the total heat loss through the fin released to the atmosphere:
\begin{equation}
\text{Total heat loss from the fin}= 4\pi R_0 \kappa B (T_w - T_a)c  \frac{K_1 (cR_0)I_1(cR_1)-I_1(cR_0)K_1(cR_1)}{K_1(cR_1)I_0(cR_0)+I_1(cR_1)K_0(cR_0)}
\end{equation}
here c is defined as $$c=\sqrt{\frac{h}{\kappa B}}$$, 2B is the thickness of the fin. Here, in Eq.(2) the $I_v(x)$ and $K_v(x)$ are the value of zero order of first kind and zero order of second kind respectively for modified Bessel function. The value of $I_v(x)$ and $K_v(x)$ is calculated online by using Keisan Casio online solver\cite{keisan}. The setup comprises of coolant flow, so we need to calculate the pressure drop in the coolant chamber and accordingly the power required by the pump to continue the flow of water. We have assumed pressure drop because of flow in straight pipe only. The drop in pressure is obtained by using the Darcy-Weisbach equation\cite{Darcy}.
\begin{equation}
\Delta P= \frac{f L\times\rho V^2}{2D_h}
\end{equation}
In fluid flow problem we need a fixed value of diameter, so we used the concept of hydraulic diameter $D_h$. It is defined an effective diameter which can replace the different geometry into a simple form.
\begin{equation}
D_h=\frac{4A}{p}
\end{equation}
In the above Eq.(3), there is a new term which is $f$ which is a Darcy friction factor. The Darcy friction factor is taken because of friction existing at the interface of coolant and solid surface. We have assumed a fully developed flow and $V$ is the velocity, which is taken as the mean flow velocity. This factor $f$ normally depends on the number of parameters like the type of flow, roughness of pipe, velocity of flow, etc. We have to make the flow as turbulent and the boundary surface as rough to remove large amount of heat from TEM sample. Accordingly, we have used the approximate solution obtained by Halland for Colabook equation\cite{Jukka}.
\begin{equation}
\frac{1}{\sqrt{f}}=-1.8\times log[({\frac{\frac{\epsilon}{D_h}}{3.7}})^{1.11}+\frac{6.9}{Re}]
\end{equation}
where $\epsilon$ is average roughness in pipe means the average height of crests and troughs, $D_h$ is hydraulic diameter, $Re$ is the Reynolds number. Using all these above mentioned formulae, we can calculate the pressure drop in pipe because of coolant flow. Now the pumping power need to supply for regular flowing of water is\cite{Murakami}
\begin{equation}
\psi = \Delta P\times V \times A
\end{equation}
where $\psi$ is pumping power required. As far as designing of TEM is concerned, our  first focus was about the length of TEM sample which should be calculated by considering the temperature dependent thermal conductivity parameter. We have to divide the entire temperature range of sample into a number of small segmented temperature difference and the difference of each portion is 5 K. The difference between each adjacent layer is assumed as small as possible to avoid the sharp variation in thermal conductivity. Accordingly, for each fixed temperature difference we have calculated the length of each individual segmented sample and finally by adding up entire length we have obtained the total length of TEM module.
\section{Calculation, Result, and Discussion}
We have briefly provided the method adopted by us in calculating the efficiency with varying the different parameters. It is clear from the above formulation that the optimization of different parameters with respect to corresponding variables for getting the efficiency will be fruitful for direct installation in automobiles. Specifically, in automobile, we have taken few experimentally obtained results which are tabulated here.\cite{Vaz}
\begin{table}[h]
\centering
\caption{Different variable parameters for a fixed prototype design}
\begin{tabular}{|p{4cm}|p{4cm}||p{4cm}|p{4cm}|}
\hline
$\text{Hot flue gas}$ & \text{Inputs} &\text{Coolant} &\text{Inputs} \\ \hline
$\text{Inlet Temperature}$ &850 K &$\text{Inlet Temperature}$ & 300 K\\ \hline
$\text{Mass flow rate}$ &30 g/s &$\text{Mass flow rate}$ &25 g/s \\ \hline
$\text{Specific Heat}$ &1063  J/kg K & $\text{Specific Heat}$ &4208 J/kg K \\ \hline
\end{tabular}
\end{table}
The mass flow rate of hot flue gas $\dot{m}= 30 g/sec$ and temperature at hot flue gas at the inlet of chamber is 850 K and temperature at the hot gas outlet is 650 K. The power dissipated by the hot flue gas when flown through the chamber using the formula 
\begin{equation}
\dot{Q}=\dot{m}\times c\times \Delta T
\end{equation}
is 6.378 kW and the energy absorbed by the coolant when passing through the chamber is 5.260 kW. Taking energy balance: \textbf{Power removed by flue gas = power gain by coolant+ power taken by circular fin + heat released to the atmosphere by convection from the unoccupied area}. So, the difference in energy (1.118 kW) will be either flown out through the circular fin or by convection to the atmosphere. The amount of energy removed by external circular extruded portion is obtained by using the Eq.(2). The value taken for convective heat transfer coefficient (h)= 100 $W/(m^2K)$, thermal conductivity of aluminium fin is 200 $W/(m.K)$ and the width of each fin is same as the width of a sample, here width is taken as 2B and 2B = 5 mm, this gives the value of c = 14.14. $R_0$= 25 mm. The value of a modified Bessel function of zero order of first kind $I_v(x)$ and second kind $K_v(x)$ is calculated and the value is shown in Table
\begin{table}[h]
\centering
\caption{Value of Modified Bessel function of first kind and second kind}
\begin{tabular}{|p{3cm}|p{2.5cm}|p{2cm}|p{2.5cm}|p{2cm}|p{2.5cm}|}
\hline
$\text{Notations}$ &$\text{Values obtained}$&$\text{Notations}$&$\text{Values obtained}$&$\text{Notations}$ &$\text{Value obtained}$ \\ \hline
$I_0(cR_0)$ first kind &1.062 & $I_1(cR_0)$&0.2551 & $I_1(cR_1)$ &0.376 \\ \hline
$K_1(cR_0)$ second kind &1.6782 &$K_0(cR_0)$ &0.9329 &$K_1(cR_1)$ &1.035\\ \hline
\end{tabular}
\end{table}
\begin{figure}[h]
\centering
\includegraphics[width=4.5in]{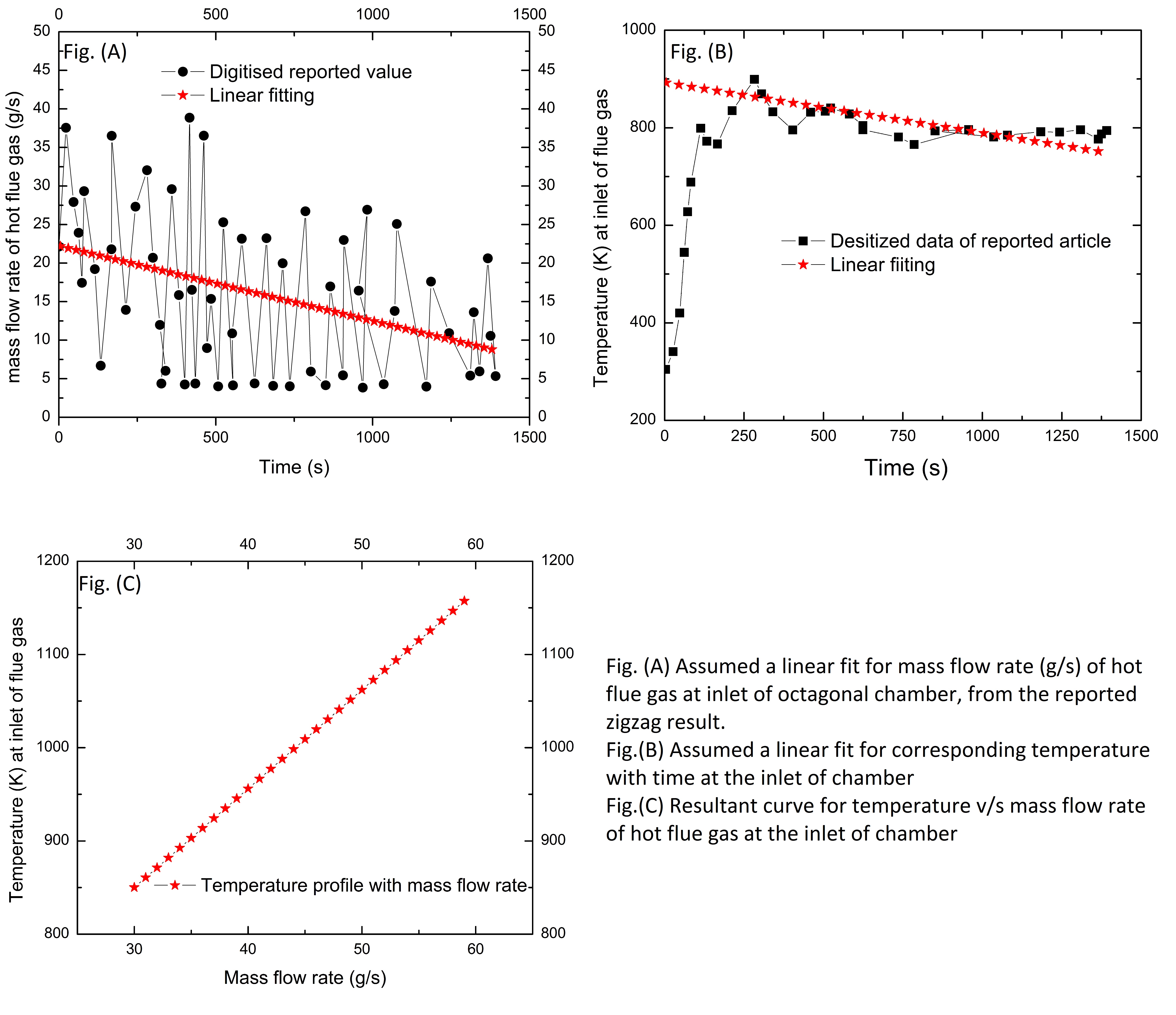}
\caption{Digitized data from the reported article and the curve we have taken for solving the problem}
\end{figure}
In the Fig.(3)(A) part is showing the reported data for mass flow rate with time, which is having zigzag curve depends on the different conditions, so we took a linear plot between mass flow rate and time\cite{Wang}. Similarly, we have temperature with time plot in Fig.(3)(B), so to make the problem simplified we have again assumed a linear curve fit\cite{Wang}. Finally, using both the linear curves we have obtained the graph between temperature and mass flow rate in Fig.(3)(C). This will enable us to obtain the efficiency by providing the temperature at the hot side of the exhaust chamber. All the following calculation is done on the basis of particular temperature range. The net calculated power passed through each fin is 24 W. Since we have assumed only five layers for modeling, so the total power removed by $fin = 24\times 5 = 120$ W. Since, we know the power extracted by convection from the lateral surface area, so using $Q = h A\Delta T$ relation we can calculate the length of hot flue exhaust gas chamber. The length calculated is 120 mm. Now the temperature distribution profile of flue gas in the exhaust pipe is assumed to be linearly varying with length. Hence, for each small TEG circular layer having thickness 5 mm, will have a temperature drop of $\sim$8 K. Using the enthalpy formula, the energy drop of flue gas when temperature drop by 8 K is 265 W. We have assumed around 10\% of power loss means 25 W drop in power because of gaps in TEM module which is filled with insulating materials between adjacent edge. So for each TEG sample at each edge of regular octagon is 240/8= 30 W. We have taken a hybridization of two different TEM for efficiency calculation and since the two materials are compatible only when their compatibility factor will match as discussed in our previous article. Considering all the theoretical aspects of the hybridization of two segments, we have to calculate the length of each segment of TEG as well as the length of the entire module. We have calculated the length of entire module by considering the variable thermal conductivity of different materials. By segmenting the entire length into a number of small segmented lengths and using the Fourier law of conduction, we have calculated the length of the sample. Segmentation technique employed is same as discussed in our earlier article\cite{Gaurav}.
\begin{figure}[h]
\centering
\includegraphics[width=3in]{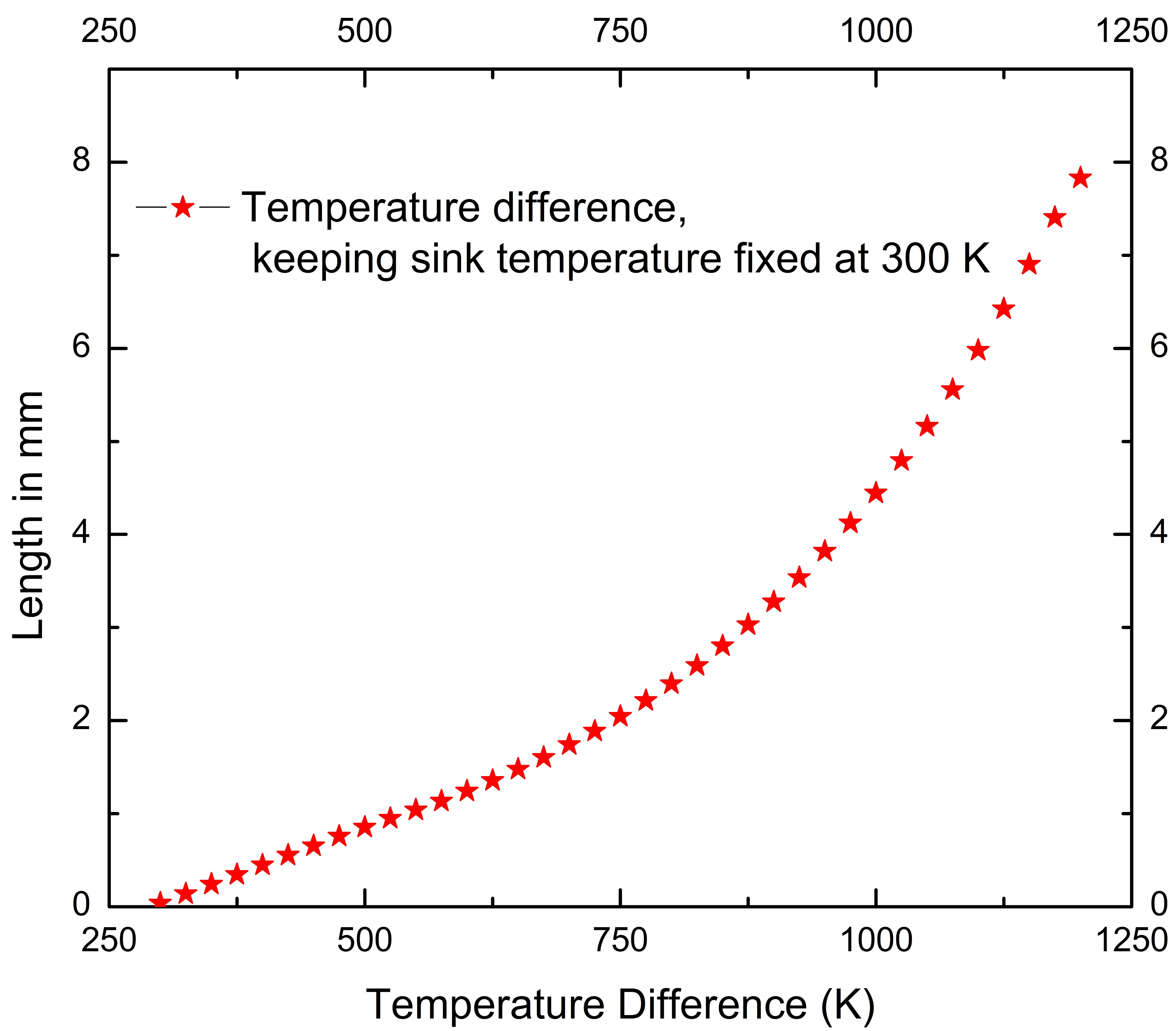}
\caption{Calculation of length of TEM sample considering temperature dependent thermal conductivity}
\end{figure}
We have to divide the entire temperature range of sample into a number of small segmented temperature difference and the difference of each portion is 5 K. The difference between each adjacent layer is assumed as small as possible to avoid the sharp variation in thermal conductivity. Considering the lowest temperature range for $Bi_2Te_3$ layer having hot end temperature is 305 K and cold end temperature is 300 K. The length for this particular layer is obtained as 0.019 mm, similarly the length of last layer means the hot end side made up of $TiO_{1.1}$ is 0.1079 mm. Accordingly, the length of different layer is obtained for a fixed temperature difference. Using all these segmented length we have calculated the entire length of sample by adding up all the segments. In Fig.(4) we have plotted variation in the length of sample with varying temperature range. This graph consist of cumulative change in length with varying the temperature range, which is very useful in selecting length of sample for any fixed temperature range. In the particular case of hybridization of $Bi_2Te_3$ and $TiO_{1.1}$ we have calculated the length as 8 mm, where length of $Bi_2Te_3$ is around $\sim$2 mm and for $TiO_{1.1}$ is around $\sim$6 mm. Another parameter is the optimization of pumping power required to maintain the pressure drop in pipe. For calculation of fluid flow through some arbitrary design, we have calculated the hydraulic diameter ($D_h$) as 3.636 mm. We took the fixed the mass flow rate of water as $\dot{m}=0.025$g/sec as reported in article and the cross sectional area as per our design. Then using the continuity equation $\dot{m}= \rho\times V\times A$ where $\rho=1000kg/m^3$ the velocity comes out to be 0.625 m/s. The Reynolds number calculated using the formula as discussed earlier is 4132, which is just above the limit of turbulent flow and as much turbulence will increase that will increase the amount of heat removal from the hot flue flow. Note:- The velocity should be around 2 m/s to avoid rusting in the pipe, but we have not achieved that value because it needs lots of power to maintain that speed but our requirement to attain turbulent flow is achieved in this mass flow rate only, so it's not economical to achieve the said speed. Now, we need to calculate the pressure drop in the pipe, but that required the value of friction factor. The value of friction factor is calculated using Halland equation is 0.06253, keeping commercially available pipe having roughness as 0.1 mm. Using the friction factor, we have calculated the pressure drop in the pipe. The pumping power requirement is obtained as 20 W.
\begin{figure}[h]
\centering
\includegraphics[width=3in]{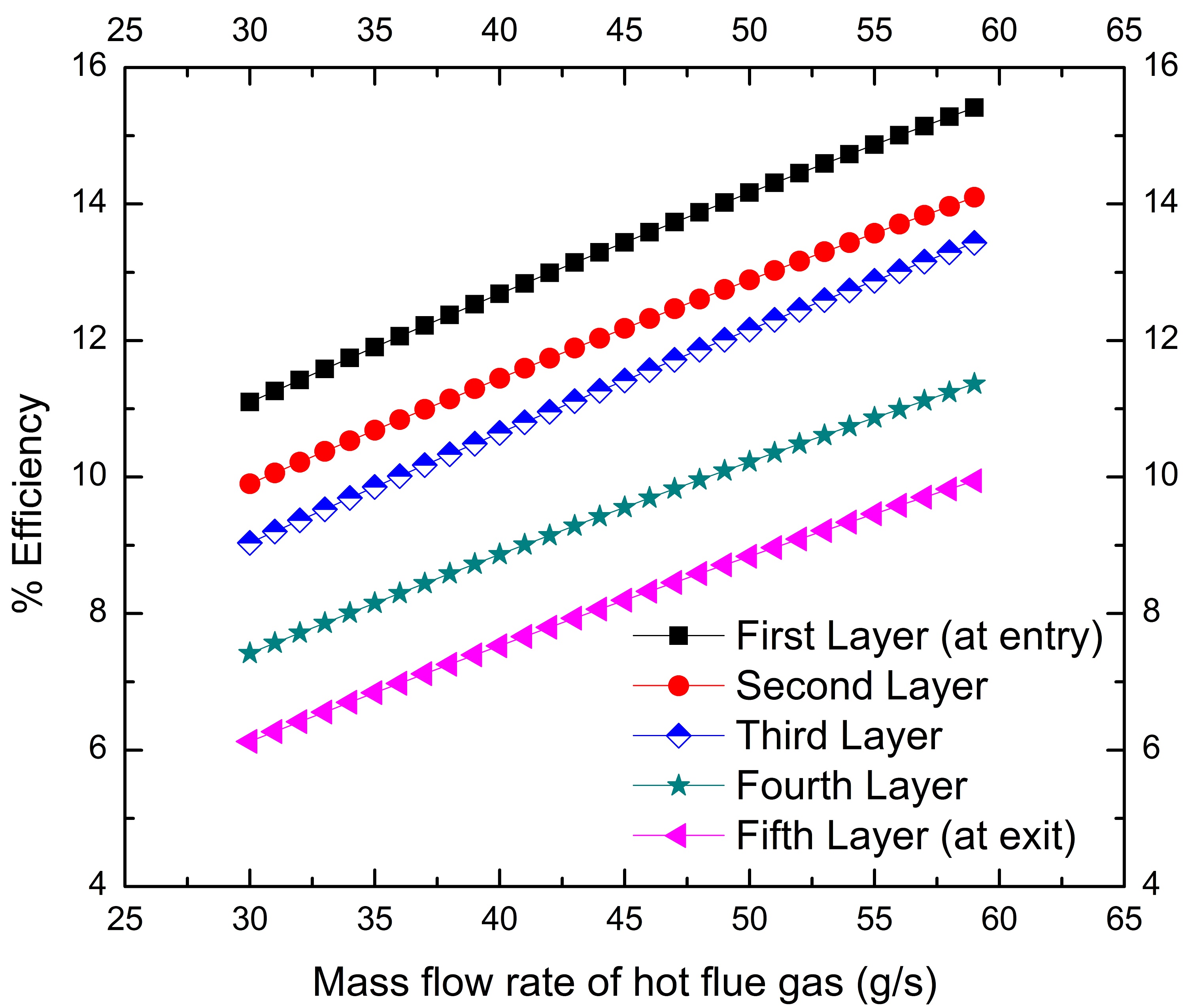}
\caption{The efficiency obtained by taking design in realistic prospects}
\end{figure}
The Fig.(5) is obtained for efficiency by installing in automobiles. Here, the plot having five curves which means for each circular layer of TEG we have different efficiency. The variation in temperature profile across the flue gas chamber is the reason behind the variation in efficiency. For varying mass flow rate of flue gas, we have calculated the temperature range of hot flue gas. Accordingly the variation in temperature of the coolant is also obtained considering the same temperature variation profile of hot gas. Using the simultaneous variation in the temperature of hot gas and coolant we have plotted the efficiency. TEM sample for TEG is taken as hybridization of two samples, namely $Bi_2 Te_3$ and $TiO_{1.1}$. The temperature of the hot and sink side is used for direct calculation of efficiency using the earlier work done by us\cite{Gaurav}. The variation in temperature and the calculation of efficiency are done by coding in Python as an open source software.
\begin{figure}
\centering
\includegraphics[width=3in]{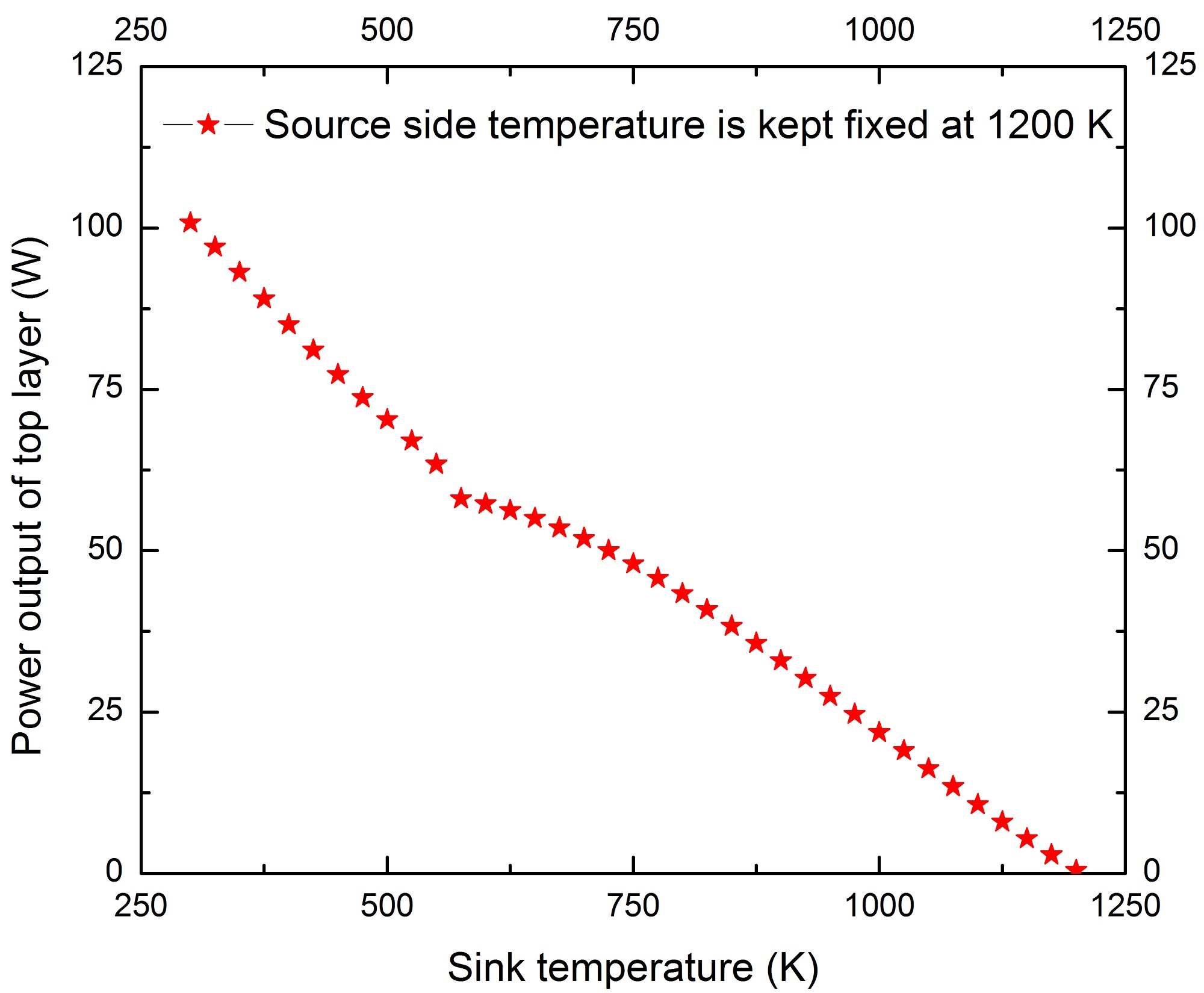}
\caption{The power output obtained by first layer of TEG, which is at the inlet of flue gas}
\end{figure}
Since, we need to find out the power output because ultimately we have to find out the net power output from our setup as done by different research groups\cite{Min, Xie, Gou}. We are concentrating on the first layer of the design. First layer is made up by the hybridization of $Bi_2 Te_3$ and $TiO_{1.1}$. Considering ideal case in which the power output by the setup is being calculated by temperature dependent materials parameter. For the calculation part we are using the formula of $Power = \frac{V^2}{R}$ where, V is potential difference generated and R is resistance of the TEM. Now, the value of V is obtained by using temperature dependent Seebeck coefficient value and R is obtained by temperature dependent electrical resistivity \cite{Okinaka, Yong} by applying the value of cross sectional area and length of sample. We have followed the same methodology of segmentation as discussed in our earlier paper \cite{Gaurav}. Interface temperature is 570 K, which is taken by considering compatibility factor. For each small segmented layer, firstly we have calculated the length of each layer for fixed temperature difference of 5 K. Each small layer will produce some power, such as the first segmented side towards the source side is producing 0.475 W. Similarly each individual segmented section will produces its own power. Since, all the segmented sections are in series so the power generated will be added up. Finally, we obtained the net power output between two ends of sample. The Fig.(6) obtained by calculating the cumulative power produced by entire segmented sections. Here, we have plotted the graph considering the variation in sink temperature of sample, where source side temperature is kept fixed with varying sink temperature. The power output of ideal case, when the hot end temperature is kept at 1200 K and sink at 300 K for the first circular layer is obtained as $\sim$ 100 W. However, for automobile case, where the source temperature remains around 800 K. So, for temperature range of 300 K to 800 K we can use same methodology as explained for fixed source temperature. For the particular case of automobile for the said temperature range was obtained as 58 W. If we consider for loss due to environment as 50\% then also the net output we will get is around 28 W. This power is obtained for the first circular layer only. Similarly using same procedure we can calculate the power output through next subsequent circular layers.\\During hybridization and installation there are problem in thermal stability. The first problem arises at the interface of two TEM and second is at the interface of sample and aluminium oxide layer. We are here trying to discuss the above said problems. As discussed earlier, we have provided the spring arrangement for longitudinal thermal variation on the surface of TEG sample and ceramic substrate. Both the segmented layers are individually examined based on their thermal expansion coefficient and found out that the expansion in the material will be counter balanced by spring arrangement. Future scope in this area is to search for the materials having thermal expansion coefficient opposite in nature for sandwiching two layers that will prevent from thermal mismatch criteria.
\section{Conclusion}
Development of technology should favour for reducing the burden over natural resources. Here we have addressed the problems regarding installation of TEG in an automobile in a simple and precise manner. Using the reported results in relation between the mass flow rate and time as well as temperature of fuel with time, we have calculated the relation between the mass flow rate and temperature. Accordingly, for the obtained temperature profile we have calculated the energy extracted to the atmosphere with different means. As far as calculation part is concerned, the heat removed by coolant is 5.260 kW during the flow of coolant through entire channel. The heat passed through each TEG module sample placed on edge is $\sim$30 W. The heat available after the coolant is removed by the fins, so the heat released by each fin is 24 W. Flue gas passes through the chamber and the length of chamber obtained is 120 mm. Turbulence in coolant flow is also optimized for extracting more heat with lesser power requirement for the pump to operate. The optimized Reynolds number of turbulent flows is 4132. Accordingly the power required to continue the flow of coolant is 20 W. Depending on the road condition the TEG should have the capability to sustain in all the condition as well the temperature of hot flue gas will change frequently. So we have plotted the efficiency curve based on the mass flow rate of hot flue gas. Power output through the first layer at the entry of TEG chamber is obtained as 100 W ideally. Their variation with sink temperature by keeping source temperature fixed, is plotted in graph. The power obtained for automobile case when temperature variation is 300 K to 800 K is 58 W. This type of result will give insight to the design engineer to think in this direction and their installation in realistic ground.

\section{Reference}
\bibliographystyle{unsrt}

\begin{thebibliography}{10}

\renewcommand{\arraystretch}{0.5}

\bibitem{Murakami}
Murakami, Yoichi, and Bora B. Mikic.
\newblock{\em IEEE Transactions on Components and Packaging Technologies} 24.1 (2001)

\bibitem{Bell}
Bell, Lon E.
\newblock{\em Science} 321.5895 (2008)

\bibitem{Yu}
Yu, Jianlin, and Hua Zhao.
\newblock{\em Journal of Power Sources} 172.1 (2007)

\bibitem{Wu}
Wu, Chih.
\newblock{\em Applied Thermal Engineering} 16.1 (1996)

\bibitem{Kraemer}
Kraemer, Daniel, et al.
\newblock{\em Solar Energy} 86.5 (2012)

\bibitem{Gaurav}
Gaurav, and Sudhir Pandey
\newblock {\em J. Renew Sustain Ener.}, {9}, 014701 (2017)

\bibitem{Chen}
Chen, Jincan.
\newblock{\em Journal of applied physics} 79.5 (1996)

\bibitem{Omer}
Omer, S. A., and D. G. Infield.
\newblock{\em Solar Energy Materials and Solar Cells} 53.1 (1998)

\bibitem{Orr}
Orr, B., et al.
\newblock{\em Applied Thermal Engineering} 101(2016).

\bibitem{Korz}
Korzhuev, M. A., and I. V. Katin.
\newblock{\em Journal of electronic materials} 39(9)(2010)

\bibitem{Liu}
Liu, X., et al.
\newblock{\em Case Studies in Thermal Engineering} 2 (2014)

\bibitem{Madhav}
Karri, Madhav A.
\newblock{\em Mechanical Engineering} 163 (2005)

\bibitem{Wang}
Wang, Yuchao, Chuanshan Dai, and Shixue Wang.
\newblock{\em Applied energy} 112 (2013)

\bibitem{Jorge}
Vázquez, Jorge, et al.
\newblock{\em Proc. 7th European Workshop on Thermoelectrics} No. 17. 2002.

\bibitem{Sumeet}
Kumar, Sumeet, et al.
\newblock{\em Journal of Electronic Materials} 44 (10) (2015)

\bibitem{Champ}
Champier, D., et al.
\newblock{\em Thermoelectric power generation from biomass cook stoves} \textbf{Energy} 35.2 (2010)

\bibitem{Champier}
Champier, D., et al.
\newblock{\em Energy} 36.3 (2011)

\bibitem{Mayer}
Mayer, P. M., and R. J. Ram.
\newblock{\em Nanoscale and Microscale Thermophysical Engineering} 10.2 (2006)

\bibitem{Yazawa}
Yazawa, Kazuaki, and Ali Shakouri.
\newblock{\em Environmental science \& technology} 45.17 (2011)

\bibitem{Lawrence}
Lawrence, E. E., and G. J. Snyder.
\newblock{\em Thermoelectrics, Proceedings ICT'02. Twenty-First International Conference on. IEEE} 2002.

\bibitem{Jahirul}
Jahirul, Mohammad I., et al.
\newblock{\em Applied Thermal Engineering} 30.14 (2010)

\bibitem{heat}
\newblock{\em www.syvum.com/cgi/online/serve.cgi/eng/heat/heat1003.html}

\bibitem{keisan}
\newblock{\em www.keisan.casio.com/2006/1180573473}

\bibitem{Darcy}
\newblock{\em www.wikipedia.org/wiki/DarcyWeisbachequation}

\bibitem{Jukka}
Kiijarvi, Jukka. 
\newblock{\em Lunowa Fluid Mechanics Paper} 110727 (2011).

\bibitem{Wang}
Wang, Chien-Chang, Chen-I. Hung, and Wei-Hsin Chen.
\newblock{\em Energy} 39.1 (2012)

\bibitem{Min}
Chen, Min, Lasse A. Rosendahl, and Thomas Condra.
\newblock{\em International Journal of Heat and Mass Transfer} 54.1 (2011)

\bibitem{Xie}
Xie, Jin, Chengkuo Lee, and Hanhua Feng.
\newblock{\em Journal of Microelectromechanical Systems} 19.2 (2010)

\bibitem{Gou}
Gou, Xiaolong, Heng Xiao, and Suwen Yang.
\newblock{\em Applied energy} 87.10 (2010)

\bibitem{Okinaka}
Okinaka, Noriyuki, and Tomohiro Akiyama.
\newblock{\em ISIJ international} 50.9 (2010)

\bibitem{Yong}
Wu, Yongjia, et al.
\newblock{\em Energy Conversion and Management} 88 (2014)






%
























%
%
%
%

\end{thebibliography}

\end{document}